# New reconstruction and data processing methods for regression and interpolation analysis of multidimensional big data

Yuri K. Shestopaloff, Alexander Y. Shestopaloff

## Abstract

The problems of computational data processing involving regression, interpolation, reconstruction and imputation for multidimensional big datasets are becoming more important these days, because of the availability of data and their widely spread usage in business, technological, scientific and other applications. The existing methods often have limitations, which either do not allow, or make it difficult to accomplish many data processing tasks. The problems usually relate to algorithm accuracy, applicability, performance (computational and algorithmic), demands for computational resources, both in terms of power and memory, and difficulty working with high dimensions. Here, we propose a new concept and introduce two methods, which use local area predictors (input data) for finding outcomes. One method uses the gradient based approach, while the second one employs an introduced family of smooth approximating functions. The new methods are free from many drawbacks of existing approaches. They are practical, have very wide range of applicability, provide high accuracy, excellent computational performance, fit for parallel computing, and very well suited for processing high dimension big data. The methods also provide multidimensional outcome, when needed. We present numerical examples of up to one hundred dimensions, and report in detail performance characteristics and various properties of new methods.

**Introduction**

Finding one or more outcome *y*-values from many input multidimensional *x*-values, often called predictors, is an algorithmically and computationally complex problem, which arises in numerous practical applications. Mathematically, such problems are addressed by different statistical and mathematical methods. Two major approaches can be distinguished in this regard. In one instance, the whole input training dataset or its large part is used to synthesize a multidimensional regression surface. A well known example can be artificial neural networks [1,2], whose different variations and more specific methods are presently used in many applications. Another example can be generalized additive models [3]. In the regression setting, its idea is to estimate the outcome *y*-value from the sum of "smooth ("nonparametric") functions", which is fit with predictors using the least squares criterion, or some more advanced method, like cubic smoothing spline or kernel smoother [3]. Despite the relative simplicity and interpretability, as the authors of [3] acknowledge, "additive models can have limitations for large data-mining applications. The backfitting algorithm fits all predictors, which is not feasible or desirable when a large number is available", so that the authors suggest using other approaches for large problems.

In the second approach, the data in a limited local area around the *x*-point, for which the outcome *y*-value to be found, are considered. The input of predictor points can be weighed based on certain considerations, like closeness of data points to the boundary of the analyzed local area. Examples can be local linear regression or local polynomial regression methods [3,4]. There are also some very custom artificial neural network methods that only look at nearby collections of points. However, to the best of our knowledge, the area of their application is very limited. The methods proposed in this work use the "local area" approach, but they allow creating multidimensional regression surfaces for datasets of virtually any size.

The idea of local regression can be understood if we recall the notion of a moving average for time series. The one-dimensional linear or polynomial kernel smoothers provide continuity and smoothness of regression lines. A similar technique can be applied for two and more dimensions. However, such applications face substantial problems. As the authors of [3] say, "While boundary effects are a problem in one-dimensional smoothing, they are a much bigger problem in two or higher dimensions, since the fraction of points on the boundary is larger. It is intuitively clear, that this fraction should grow fast with the increase of dimensionality. Indeed, the authors in [3] acknowledge that "In fact, … the fraction of points close to the boundary increases to one as the dimension grows." Since the idea of the method is that boundary points should contribute less, in order to provide smoothness of the approximating function, such an increase poses a problem. The local polynomial regression is of help in such situations. However, as the authors of [3] say, "Local regression becomes less useful in dimensions much higher than two or three." The reason is the contradiction between the requirements of maintaining low bias (localness) and having a "sizable sample in the neighborhood" in order to preserve the low variance with the increase of dimensions. For that, the total sample size has to increase exponentially, which in turn, even if the requirement is fulfilled, brings many other problems. So, by and large, the method does not fit multidimensional problems.

The methods, which we propose, work with very high dimensional data, like neural networks do, but require much less - by orders of magnitude - computational resources, both in terms of processing power and memory. This largely comes for the price that our methods do not support some features, which artificial neural networks do. First of all, this is a simultaneous incorporation of all training points (which is not needed for many tasks). On the other hand, our methods have certain advantages the artificial neural networks do not offer.

## Methods and Results

### Gradient based method

In this section, we introduce a method for finding the outcome *y*-values based on *x*-predictors, which can be used for many purposes, like interpolation, restoring and filling data in multidimensional datasets. The method uses outcome values (*y*-values) of nearby points (local area approach). It is based on the usage of gradients, and so it is called as a gradient based method below. A note should be made that this gradient based method has little in common with the well known gradient methods such as the gradient descent (the original method can be found in [5], while review of later developments in [6]). So, despite the same word, the new method assumes different than conventional connotation.

The idea of the method is illustrated by Fig. 1 for a 3-D space. The point $Y_k$, for which we want to find a value of $y_k = f(X_k)$ is surrounded by points $\{Y_1, Y_2, Y_3, Y_4\}$.

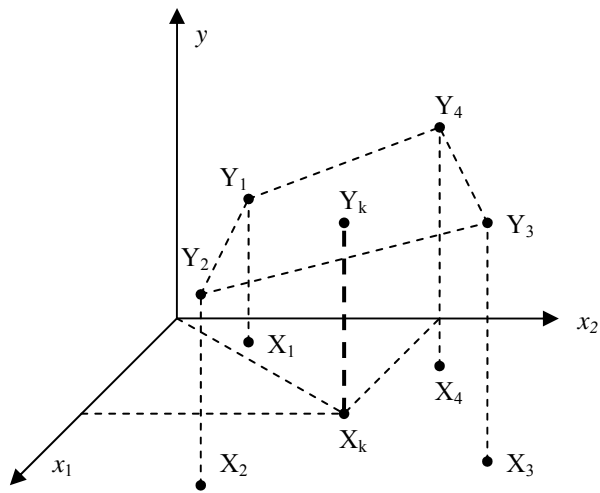

Fig. 1. Finding an approximate value of $y_k = f(X_k)$ based on values of known neighboring points $\{Y_1, Y_2, Y_3, Y_4\}$.

Three points $\{Y_1, Y_2, Y_3\}$ define a plane (it is known from geometry that if three points are not on the same line, an unambiguously defined plane always exists that goes through these points). Since the plane is described by a linear equation $y = ax_1 + bx_2 + c$, where $a$, $b$, $c$ are constant, the partial derivatives of $y$ for any point on this plane are the same. Similarly, we can consider a plane in a space with arbitrary dimension $N$, in which $N$ points not belonging to the same multidimensional line unambiguously define a $N$-dimensional hyperplane. The partial derivatives for any point of this hyperplane, similarly to a 3-D space, will be also the same. Knowing these partial derivatives, we can always find a distance between two points on such a hyperplane, and accordingly projections of this distance on the coordinate axes, including the $y$-axis. This is the core idea behind the first proposed method.

Suppose, we have a dataset $\{y_i\}$, such that

$$y_i = f(x_1^i, x_2^i, \ldots x_n^i) \tag{1}$$

where $f$ is a function of $n$ variable parameters, and the dataset $X_i = (x_1^i, x_2^i,,...x_n^i)$, $i = 1,2,...I$, containing $I$ points, represents a continuum of points in $n$-dimensional space. The appropriate $(n+1)$ dimensional dataset that includes $y$ is denoted as $Y_i = (x_1^i, x_2^i,,...x_n^i, y^i)$. Thus, in geometrical terms, $\{y_i\}$ represents points on a surface in a $(n+1)$ dimensional space. The dataset $T_i = (x_1^i, x_2^i,,...x_n^i, y^i)$ is our training dataset. The dataset $\{X_k\}$, $k = I+1,...K$, represents a continuum of points we would like to find the values $\{y_k\}$ for.

The differential of value of $y = f(X)$ is as follows.

$$dy = \frac{\partial f}{\partial x_1} dx_1 + \frac{\partial f}{\partial x_2} dx_2 + ... + \frac{\partial f}{\partial x_n} dx_n \tag{2}$$

So, once we know, let us say, the value of $y_1 = f(X_1)$ at a reference point $Y_1$, we can find an approximate value of $y_k = f(X_k)$ as follows.

$$y_k = f(X_k) \approx f(X_1) + \frac{\partial f(X_1)}{\partial x_1}(x_1^k - x_1^1) + \frac{\partial f(X_1)}{\partial x_2}(x_2^k - x_2^1) \tag{3}$$

In Eqn 3, we do not know the values of partial derivatives, whose approximate values can be found from the system of appropriate equations using the training dataset. Let us denote for convenience $p_j = \frac{\partial f(X_1)}{\partial x_j}$, $j = 1,2$. Then, using two training points $Y_2$ and $Y_3$, we can write the following system of equations.

$$p_1(X_1)(x_1^2 - x_1^1) + p_2(X_1)(x_2^2 - x_2^1) = y_2 - y_1 \tag{4}$$

$$p_1(X_1)(x_1^3 - x_1^1) + p_2(X_1)(x_2^3 - x_2^1) = y_3 - y_1 \tag{5}$$

Solving this system of equations, we obtain the values of unknown partial derivatives $p_1$ and $p_2$. Substituting them into Eqn 3, we will find the value of $y_k$.

When the training points are located in the same plane $(x_1, y)$ or $(x_2, y)$, then, accordingly, $(x_2^2 - x_2^1) = 0$ or $(x_1^3 - x_1^1) = 0$, and the partial derivatives can be found as follows.

$$p_1(X_1) = (y_2 - y)/(x_1^2 - x_1^1)$$

$$p_2(X_1) = (y_3 - y_1)/(x_2^3 - x_2^1)$$

Note that we can use any combination of three points from four presented in Fig. 1 (that is four combinations in total), like the training dataset $\{Y_1, Y_3, Y_4\}$ instead of $\{Y_1, Y_2, Y_3\}$. (The choice should be based on the closeness of the approximated point to the training points and other possible considerations related to the problem. In our case, judging by the drawing in Fig. 1, this is the dataset $\{Y_1, Y_3, Y_4\}$.) Accordingly, for our example, we will have four pairs of partial derivatives and four values of $y_k$. Their average value, in general, provides a more accurate estimation.

So, we found the following:

(a) The gradient based method requires minimum number of training points equal to (n+1), that is to the dimension of space;

(b) Data redundancy can be used to increase the accuracy of approximation.

The generalization of Eqns 3-5 is as follows.

$$y_k = y_r + \sum_{i=1}^{i=n} \frac{\partial f(X_r)}{\partial x_i}(x_i^k - x_i^r) \qquad (6)$$

where $y_r$ corresponds to the reference point $Y_r$ (which is $Y_1$ in our example).

The system of $n$ equations for finding $n$ unknown values of $p_j$, $j = 1, 2, \ldots n$, becomes as follows.

$$\sum_{i=1}^{i=n} p_i(X_r)(x_i^2 - x_i^1) = y_2 - y_1$$

$$\sum_{i=1}^{i=n} p_i(X_r)(x_i^3 - x_i^1) = y_3 - y_1$$

$$\ldots$$

$$\sum_{i=1}^{i=n} p_i(X_r)(x_i^{n+1} - x_i^1) = y_{n+1} - y_1 \qquad (7)$$

Here, we assumed that the reference point is $Y_1$.

### Numerical example

Let us consider the following function of three variables.

$$y = x_1^3 + 0.4\sin(6x_2) + 0.6\sin(4x_3 + 0.5) \qquad (8)$$

The cross-sections of this 4-D function along the (*x, y*) planes are shown in Fig. 2. All three variables in Fig. 2 change in the intervals from 0 to 3, with the number of subintervals for each variable of 30. We can see that this 4-D surface has very uneven surface with frequent, irregularly located, "hills", "hollows" and troughs.

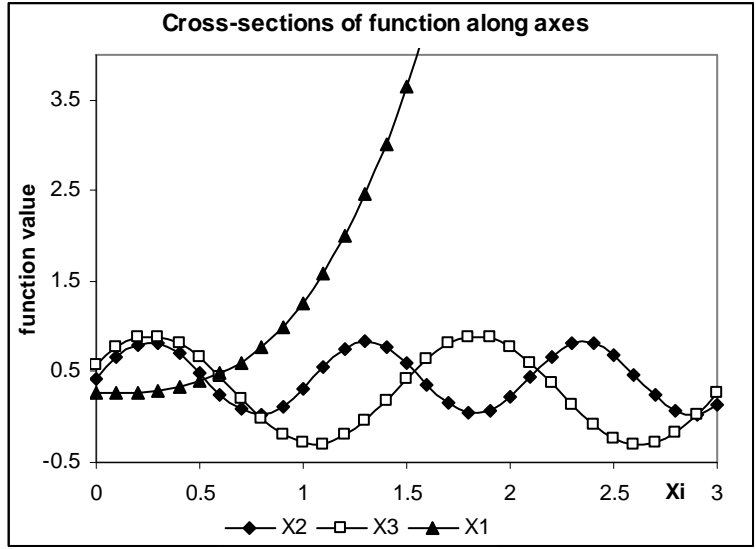

Fig. 2. Cross-sections of the approximated function along the coordinate axes.

Results of computations are presented in Table 1. We computed the values of *y* for the reconstructed points located between training points (at a relative distance - compared to the discrete intervals - approximately from 0.3 to 1/2 from the reference point along each coordinate, so that no points lay in the same 2-D coordinate plane ). Algorithms were implemented in a C++ single threaded application with STL (Standard Template Library). Relative error was estimated with regard to the average difference in *y* values corresponding to the reconstructed point and to the nearby reference point. Calculation time is provided for a laptop computer with a 2-core 2 GHz Intel processor.

    The training points were generated on a 4-D rectangular mesh, whose coordinate values could be modulated by small noise, not exceeding the fraction of mesh interval. We used the noise option for testing the procedure for finding gradients using the system of linear equations Eqn 7, while the presented statistical valuations were done without noise for *x*-coordinates. The effect of noise for *y*-values was considered separately. The system of linear equations was solved by Gauss method.

In Table 1, the column "Average Y differ." shows the average absolute difference between the *y*-values of the reference and the reconstructed points. The column "Relat. error for $Y_{comp}$" shows the ratio of the absolute error for $Y_{comp}$ (meaning, of course, the average value) and the value of "Average Y differ.", from the second column.

Table 1. Reconstruction of data points in between 4-D training points.

| Scenario No. of points | Average Y differ. | Abs. error for $Y_{comp}$ | Max. abs. error $Y_{comp}$ | Relat. error for $Y_{comp}$ | Calculation time, sec |
|---|---|---|---|---|---|
| 8000 | 0.428 | 0.0338 | 0.105 | 0.079 | 0.047 |
| 24389 | 0.265 | 0.0146 | 0.047 | 0.0551 | 0.11 |
| 93639 | 0.166 | 0.00691 | 0.0206 | 0.0418 | 0.44 |
| 790000 | 0.087 | 0.00175 | 0.0052 | 0.0203 | 3.82 |
| 2664120 | 0.057 | 0.00078 | 0.0023 | 0.0137 | 12.7 |

As we can see from Table 1, the accuracy of calculated *y*-value increases with the decrease of discrete intervals, which was expected. In all presented scenarios, the accuracy is reasonably good, ranging from 1.37% for 2,664,120 points to 7.9% for 8000 points, relative to the difference between *y*-values, corresponding to the reference and reconstructed points. Note the complex shape of the surface we considered. Such accuracy is acceptable in many multidimensional regression problems. So, the introduced gradient based method can be considered as a practical one, suited for many real applications.

Fig. 3 shows dependence of accuracy and calculation time on the number of training points. If time is drawn in a logarithmic scale, then the dependence is close to a linear one. In other words, calculation time is about proportional to the number of computed points, which could be expected.

Note the decrease of the absolute error by 43 times versus the increase of the number of points by 333 times. We should not expect linear dependence in this case, since what matters more is the length of subintervals, which decreased by about 7.5 times. This value is on par with the increase of relative accuracy (by 5.77 times).

Note that this method can be used when coordinates $\{x_i\}$, which are dependent or independent. Such a feature is important for practical applications when dependence information is not available, which is often the case.

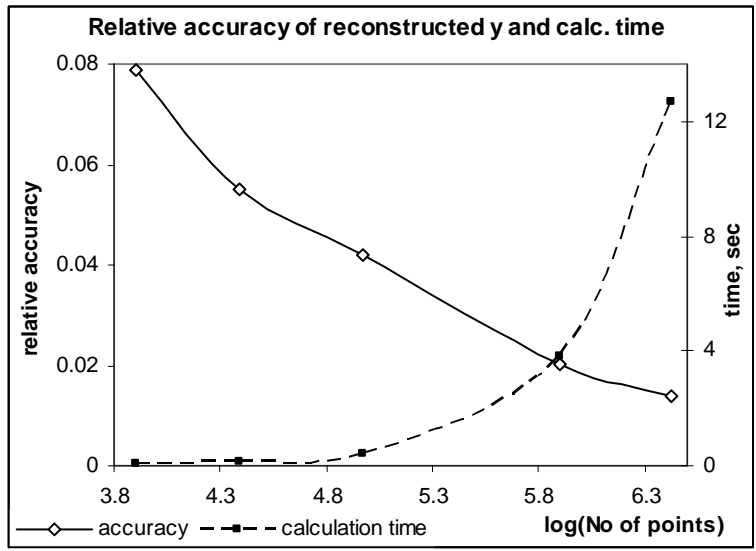

Fig. 3. Dependence of relative accuracy of computed *y*-values, and the calculation time on the logarithm of number of points.

**Approximating "smooth" multidimensional surfaces**

*The concept of adjusting gradient values*

The nature of many phenomena of practical interest, which can be described by multidimensional surfaces like Eqn 1, is usually continuous. This continuity may assume not only the continuity of the function itself, but also the continuity of at least first derivatives. In this regard, the gradient based method introduced above preserves the continuity of a multidimensional function, but not the continuity of its first derivative. The last requirement can be addressed by the following method proposed below.

This method can be considered as a qualitative enhancement of the gradient based method. We first find the unknown *y*-values, using the hyper-plane, and corresponding gradients along the *x*-coordinates, then adjust the values of gradients and calculate the unknown *y*-value using Eqn 3. One can think of this procedure as rotating the original hyper plane, obtained by the gradient based method, around the reference point, in order to more accurately match the hyperplane with the real surface.

The main idea of the method is to approximate not the entire surface, but its cross-sections by 2-D planes $(x_i, y)$. Such cross-sections are approximated by smooth curves satisfying certain requirements. Then, we find the point of intersection of such an approximating curve and the projection of line $X_k Y_k$ (see Fig. 1) on the appropriate $(x_i, y)$ plane. Knowing this intersection point, we can find the value of adjustment for the gradient.

For the method explanation, let us consider an example of 3-D space in Fig. 4. The solid curve represents the intersection of the 3-D surface $y = f(x_1, x_2)$ with the plane $(x_2, y)$. For convenience of presentation, we assume that all points $\{X_i\}, i = 0,1,2,3$, are located on the axis $x_2$, although the same consideration are valid for any plane parallel to the plane $(x_2, y)$, and the solid curve goes through the points $\{Y_i\}, i = 0,1,2,3$. Tangential lines to this curve at the points $Y_1$ and $Y_2$ are accordingly $t_1$ and $t_2$. The y-value computed by the gradient based method is represented by a point $Y_{CG}$. The y-value, found by a new method, is represented by the point $Y_{CM}$.

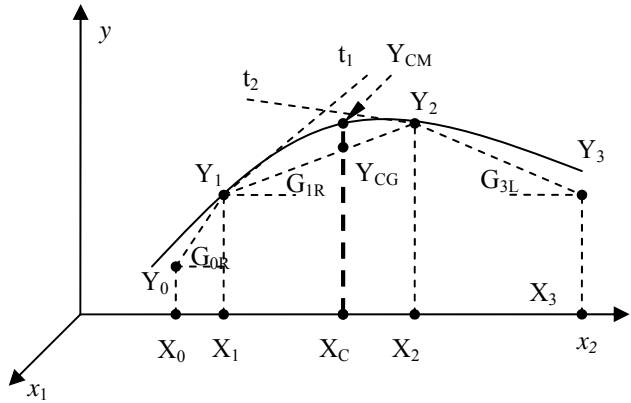

Fig. 4. Finding y-value using the smooth approximating surface. Calculating an adjustment value for the gradient in the $(x_2, y)$ plane.

Our goal is to find its y-value. Gradients for the straight lines $Y_0 Y_1$, $Y_1 Y_2$, $Y_3 Y_2$ are accordingly $G_{0R}$, $G_{1R}$ and $G_{3L}$ (index 'R' denotes the gradient to the right and 'L' to the left). For the line $Y_2 Y_3$ (note that the direction of line matters!) the gradient $G_{2L} = -G_{1R}$

(in other words, the left and right gradients for the same line have opposite algebraic signs.)

Note that the gradient $G_{0R}$ *is not equal* to the slope of the tangential line $t_1$, since the last one is tangential to the approximation curve; it is not a continuation of the straight line $Y_0 Y_1$. Similarly, the tangential line $t_2$ is not a continuation of the straight line $Y_3 Y_2$. Finding the gradients for these tangential lines is a separate task, which will be considered later.

Now, let us approximate the curved solid line between points $Y_1$ and $Y_2$. We consider first the horizontal orientation of the line $Y_1 Y_2$, and then make adjustment for the gradient $G_{1R}$. Fig. 5 presents this fraction of the approximation curve in detail. (Note that the line *OB* corresponds to line $Y_1 Y_2$ in Fig. 4.)

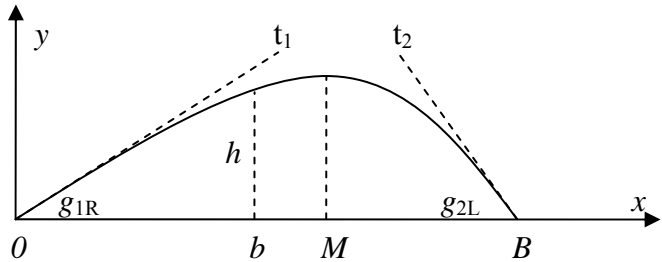

Fig. 5. Defining the approximation function based on values of gradients.

The criteria, to which the approximation curve should satisfy, are defined as follows.
(a) Its first derivatives at points *O* and *B* have to be equal to gradients $g_{1R}$ and $g_{2L}$ accordingly.
(b) The maximum (point *M* in Fig. 5) should be displaced towards the end with a greater gradient.

The following function of argument *x*, which is a cubic polynomial, satisfies these criteria.

$$y = Kx(B-x)(g_{1R}(B-x) + g_{2L}x) \tag{9}$$

Its introduction is based on a consideration that the height $h$ should be equal to zero at the ends of the interval (0, $B$), and that

$$\lim_{b \to 0} h/b = g_{1R} \tag{10}$$

$$\lim_{b \to B} h/(B-b) = g_{2L} \tag{11}$$

Substituting Eqn 9 into Eqns 10 and 11, we find $K = 1/B^2$ in both cases. The graphs of this function, when gradients that have the same and the opposite algebraic signs, are shown in Fig. 6.

Obtaining the same value of scaling coefficient of $K = 1/B^2$ for both ends of the interval is of great convenience. If we were not able to do so, the situation would not be hopeless, of course, but the things would become more complicated.

Note that the family of functions in the form

$$y = Kx(B-x)(g_{1R}(B-x)^d + g_{2L}x^d) \tag{12}$$

where $d > 0$, $K = 1/B^{k+1}$, also preserve the gradients at the ends of the interval. However, when $d > 1$, these functions acquire inflection points, and, in certain conditions, such inflection points can appear within the interval (0, $B$), which we use for approximation. With the further increase of $d$ above one, this function becomes bimodal, which makes it rather useless for approximation purposes, unless there are some specific requirements.

When $d \to 0$, the functions are unimodal, and are very similar in appearance to functions shown in Fig. 6, which makes such functions potentially good candidates for approximation purposes too. Some problem might present the fact that their maximums then begin to shift towards the middle of the interval, which still could be acceptable in some applications.

The approximation function has to be smooth at the training points (having continuous first derivative), on one hand, and it also has to have the ability to change its amplitude. Functions (9) and (12) satisfy these requirements. Apparently, other families of functions with similar properties can be used, depending on a particular problem.

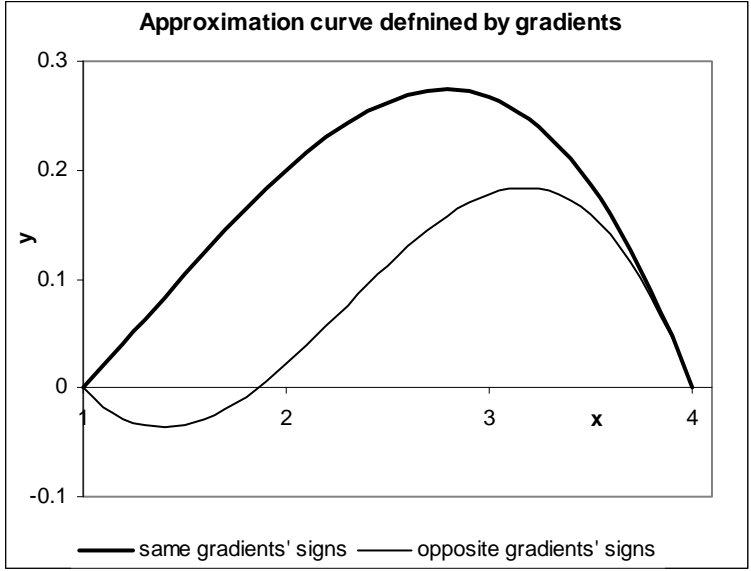

Fig. 6. Approximation function with the same and opposite signs of gradients at the ends of the interval.

### *Matching first derivatives at turning points. Gradient adjusting*

Fig. 7 shows how to find the angles $F_{g1}$, corresponding to gradient $g_{1R}$ from Fig. 5. Since we know the angles $F_0$ and $F_1$, corresponding to gradients for the lines $Y_0Y_1$ and $Y_1Y_2$, we can find $F_{g1}$ as

$$F_{g1} = -(F_1 - F_0)/2 \tag{13}$$

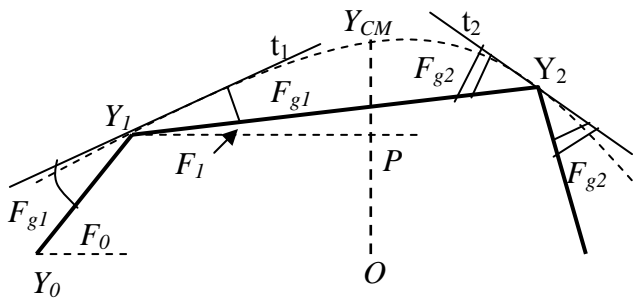

Fig. 7. Finding angles corresponding to gradients for approximation function.

Similarly, we can find the angles $F_{g2}$. Note that the angles $F_{g1}$ (angles from left and right of the point $Y_1$) between the tangent $t_1$ and lines $Y_0Y_1$ and $Y_1Y_2$ are equal, which preserve the continuity of the first derivative of approximating functions from left and right of the point $Y_1$.

As it was said before, the main idea behind exploiting approximation functions is using the adjusted values of gradients. Such, for the line $Y_1Y_2$, the gradient is equal to $g_1 = \tan(F_1)$, Fig. 7. However, for finding the value of $Y_{CM}$, we will use the adjusted gradient (Fig. 8).

$$g_{1cor} = \tan(F_1 - F_{2C}) \qquad (14)$$

Here, $F_{2C}$ is the angle between the base line $OB$ (corresponding to line $Y_1Y_2$ in Fig. 7) and the line $BY_{CM}$ in Fig. 8. The distance $OB = Y_1Y_2$ can be found as the difference between the appropriate $x$-coordinates of points $Y_1$ and $Y_2$ divided by cosine of the angle $F_1$: $OB = (x^{(2)} - x^{(1)})/\cos F_1$, where upper indexes in brackets denote $Y$ points' numbers.

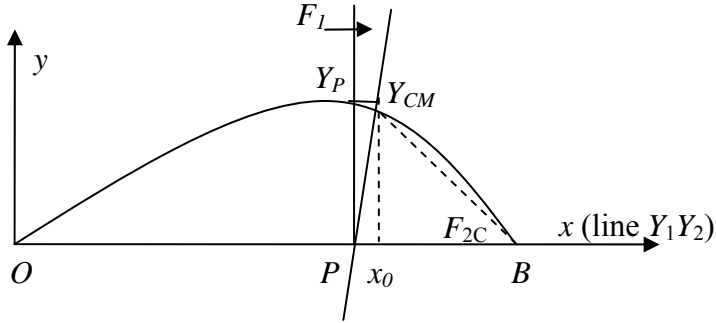

Fig. 8. Finding the adjustment to a gradient's value using approximation function.

Note that we do all calculations in the new system of coordinates, transformed from the system of coordinates corresponding to Fig. 7 by rotation by angle $F_1$ counterclockwise, so that the system of coordinates $xy$ in Fig. 8 is actually the system of coordinates, in which Eqns 9 and 12 are presented.

We can find the point $Y_{CM}$ as a point of intersection of the approximation function $A(x)$, defined by Eqn 9, and the line $PY_{CM}$ by solving the following equation.

$$x(B-x)(g_{1R}(B-x) + g_{2L}x)/B^2 = kx + c \qquad (15)$$

Here, the right part of Eqn 15 represents the line $PY_{CM}$: $k = 1/\tan F_1$, $c = -kx_P$, where $x_P$ is the *x*-coordinate of point *P*. Although the cubic equation Eqn 15 can be solved analytically, it is probably better solving it numerically in order to avoid an ambiguity with multiple roots. In our case, we have a good first approximation point $x_0$ for the iterative solution procedure, which is defined as follows (Fig. 8).

$$x_0 = x_P + y_p \tan(F_1) \tag{16}$$

The Newton-Raphson's iterative method works well for finding solution of Eqn 15. In our calculations, even for the high accuracy of $10^{-9}$ (meaning the difference between successive approximations), we did not need to use more than three iterations, which is largely due to a good first approximation value defined by Eqn 16, and the smoothness of approximation function defined by Eqn 9. We considered approximating functions with one and two extremums (see Fig. 6), and both concave and convex (in case of one extremum).

*Numerical example*

Similarly to the gradient based method, we calculated unknown *y*-values for different scenarios, and compared the obtained accuracy of approximation with the gradient based method.

We considered the following three test functions representing 4-D concave and convex surfaces, and also the same highly irregular 4-D surface defined by Eqn 7, which we used for the study of the gradient based method.

1. $y = 0.3x_1^{0.5} + 0.5x_2^{0.5} + 0.7x_3^{0.5}$
2. $y = 0.3x_1^{1.3} + 0.5x_2^{1.5} + 0.7x_3^{1.8}$
3. $y = x_1^3 + 0.4\sin(6x_2) + 0.6\sin(4x_3 + 0.5)$

The results are shown in Table 2. Simultaneously, for comparison, we presented the *y*-values by the gradient based method too. The column 4, similar to analogous column in Table 1, shows the ratio of the average absolute error of computed *y*-values and of the average difference between the *y*-values of the reference and reconstructed points from column 2.

Table 2. Reconstruction of data points in between 4-D training points using smooth approximation surface.

| 1. No. points | 2. Average differ. betw. the reference and computed y | | | 3. Average abs. error for computed y | | | 4. Error for computed y relative to differ. betw. refr. and calc. points | | | 5. Error relative to gradient based method | | | Calc time sec |
|---|---|---|---|---|---|---|---|---|---|---|---|---|---|
| Function | 1 | 2 | 3 | 1 | 2 | 3 | 1 | 2 | 3 | 1 | 2 | 3 | |
| 8000 | 0.03 | 0.29 | 0.98 | $1.05 \times 10^{-6}$ | $6.3 \times 10^{-6}$ | 0.017 | $3.3 \times 10^{-5}$ | $2.1 \times 10^{-5}$ | $1.7 \times 10^{-2}$ | $3.1 \times 10^{-3}$ | $2 \times 10^{-3}$ | 0.43 | 0.18 |
| 117649 | 0.013 | 0.11 | 0.37 | $8.8 \times 10^{-8}$ | $1.4 \times 10^{-7}$ | $5.8 \times 10^{-4}$ | $6.9 \times 10^{-6}$ | $1.2 \times 10^{-6}$ | $1.5 \times 10^{-3}$ | $1.6 \times 10^{-3}$ | $2.6 \times 10^{-4}$ | 0.09 | 2.8 |
| 1000000 | 0.006 | 0.06 | 0.19 | $1.3 \times 10^{-8}$ | $1.9 \times 10^{-8}$ | $4 \times 10^{-5}$ | $2 \times 10^{-6}$ | $3.4 \times 10^{-7}$ | $2.1 \times 10^{-4}$ | $9 \times 10^{-4}$ | $1.4 \times 10^{-4}$ | 0.03 | 22 |

We did not present the maximum error, like in Table 1. It did not differ significantly from the average absolute error, exceeding it not more than 2 - 3.5 times.

As we can see from Table 2, the proposed method with smooth approximation surface is far more superior to the gradient based method, although, as we earlier discussed, the gradient based method is also of practical value for many problems. The accuracy the new method provides (column 5) is better from tens of times to several thousand times than the accuracy of the gradient based method for the same data. The only exception is when we have a highly irregular surface, 3, with large discrete intervals. However, even in this case, the accuracy is 2.3 times better than the gradient based method produces. Both absolute and relative accuracies are very good. So, overall, this method is a very substantial advancement compared to the gradient based method.

The computational performance should be also considered as very good and, depending on the problem, in many instances as exceptional. The reason is that unlike in many other methods, which have to create the entire multidimensional surface for all training data first (which takes lots of time and computational resources, like in case of artificial neural networks), the proposed method does not need creating an entire multidimensional surface, but recreates only part of the total surface using data nearby the computed data point. For the modern computers and typical data streams, it means real time or close to real time performance.

**Accuracy and computational performance of algorithms in high-dimensional spaces**

The proposed algorithms are easily scalable for high dimensions. We calculated algorithms' performance characteristics for the following three functions using the same laptop computer. Table 3 shows results for a concave function 1 and a convex function 2. Results for the irregular function 3 are close to the presented values (not shown). Below, $N$ is the total number of space dimensions (including $y$-coordinate)

1. $y = \sum_{i=0}^{i=N-2}(0.3+i/(4(N-1)))x_i^{0.5}$

2. $y = \sum_{i=0}^{i=N-2}(0.3+i/(4(N-1)))x_i^{1.5}$

3. $y = x_0^{1.5} + \sum_{i=1}^{i=N-2}\sin\left(x_i\left(0.4+\frac{i}{2(N-1)}\right)\right)$

Table 3. Accuracy and computational performance characteristics of the method with smooth approximation surface, versus the gradient based method depending on the dimensionality of space.

| 1. Dimension of of $x$ space | 2. Average differ. betw. the reference and computed y | | 3. Abs. error for computed y | | 4. Error for computed y relative to differ. betw. the refr. and calc. points | | 5. Accuracy relative to the gradient based method | | 6. Calc. time per point for the gradient based method and for the smooth surface method, in *sec* | |
|---|---|---|---|---|---|---|---|---|---|---|
| Func. | 1 | 2 | 1 | 2 | 1 | 2 | 1 | 2 | Grad | Smooth |
| 10 | 0.13 | 0.4 | $1.3\times10^{-5}$ | $4.6\times10^{-6}$ | $1.0\times10^{-4}$ | $1.1\times10^{-5}$ | $4.7\times10^{-3}$ | $5.5\times10^{-4}$ | $3\times10^{-5}$ | $1\times10^{-4} \div 3.6\times10^{-4}$ |
| 30 | 0.42 | 1.21 | $7.1\times10^{-5}$ | $4.3\times10^{-5}$ | $1.7\times10^{-4}$ | $3.6\times10^{-5}$ | $7.6\times10^{-3}$ | $1.6\times10^{-3}$ | $1.7\times10^{-4}$ | $7\times10^{-4} \div 7.4\times10^{-3}$ |
| 50 | 0.7 | 2.04 | $1.2\times10^{-4}$ | $7.8\times10^{-5}$ | $1.7\times10^{-4}$ | $3.8\times10^{-5}$ | $7.7\times10^{-3}$ | $1.7\times10^{-3}$ | $4.2\times10^{-4}$ | $2\times10^{-3} \div 3.2\times10^{-2}$ |
| 100 | 1.4 | 4.1 | $2.5\times10^{-4}$ | $1.6\times10^{-4}$ | $1.8\times10^{-4}$ | $3.9\times10^{-5}$ | $7.8\times10^{-3}$ | $1.8\times10^{-3}$ | $1.7\times10^{-3}$ | $7\times10^{-3} \div 2.3\times10^{-1}$ |

The column 4, similar to analogous column in Table 2, shows the ratio of the average absolute error for computed $y$-values, and of the average difference between the $y$-values of the reference and reconstructed points from column 2.

We can see from Table 3 that the accuracy (column 3) depends on the shape of approximated function (in our case, concave or convex). In particular, the accuracy is several times better for a convex function (the difference also depends on where to choose the reference point - on the right or on the left, meaning values of *x*-coordinates for the unknown and reference points).

The relative error (column 4) weakly depends on the dimensionality, increasing by 1.8 - 3.5 times when the total number of dimensions (with the addition of *y*-coordinate) changes from 11 to 101. Comparison with the gradient based method (column 5) shows that the accuracy of the method with smooth approximation surface is by far superior (from 130 to 1820 times in our example).

***Influence of y-errors on the accuracy of approximation***

We also considered scenarios when the input *y*-values were estimated with errors. For the error generation, we used the normal distribution with a zero mean and the values of standard deviations from 0.01 to 0.3. For the original surface, we used functions 1 and 2 from the previous section. The dependence of accuracy on the number of dimensions and the value of standard deviation was estimated by two ratios. The first ($R_1$) is the ratio of the absolute value of the average deviation of *y*-values from the original surface, to the absolute value of the average deviation of computed *y*-values from the original surface.

$$R_1 = \frac{\sum_{i=1}^{M} \left| y_i - y_i^{orig} \right|}{\sum_{i=1}^{M} \left| y_i^{comp} - y_i^{orig} \right|} \tag{17}$$

Here, the index '*orig*' corresponds to original values defined by functions 1 or 2, the index '*comp*' - to computed *y*-values.

The second ratio ($R_2$) is similar, only instead of the absolute value of the average deviation of computed *y*-values from the original surface we use their algebraic sum.

$$R_2 = \frac{\sum_{i=1}^{M} \left| y_i - y_i^{orig} \right|}{\sum_{i=1}^{M} (y_i^{comp} - y_i^{orig})} \tag{18}$$

We did not discover any meaningful dependence of the ratios on the value of standard deviation and type of function. On the other hand, the ratio depends on the number of dimensions. Namely, it is about inversely proportional to the number of dimensions. Table 4 presents the results.

Table 4. Dependence of the method's accuracy on the number of dimensions.

| Number of dimensions | 10 | 30 | 50 | 100 |
|---|---|---|---|---|
| Ratio $R_1$ | 0.31 | 0.11 | 0.058 | 0.032 |
| Ratio $R_2$ | 1.0 | 0.7 | 0.4 | 0.17 |

We also repeated the same calculations for even distribution within the range from 0.05 to 0.3. Results were similar.

***Improving methods' accuracy by averaging y-values. Overfitting issue***

Dependence of accuracy on the number of dimensions presented in Table 4 can be explained as follows. Let us look at Eqn 6. Since we use the absolute values in Eqn 17, it means that the average absolute error per reconstructed point is proportional to the number of dimensions, while the average displacement of initial *y*-values from the original surface remains the same regardless of the number of dimensions. This is what we see in Table 4 for the ratio $R_1$. On the other hand, when we use the algebraic sum from Eqn 18, the terms in the sum in Eqn 6 can be both positive and negative, which accordingly reduces the denominator and provides greater values of $R_2$. Note that this parameter experiences large fluctuations, depending on particular distribution, for dimensions less than 30 (up to 6-7 times in our calculations), while the value for 100 dimensions is stable. In principle, the ratio $R_2$ should not noticeably decrease with the dimension growth. There are several plausible factors, we can think of, which could provide such an effect. However, which particular factors decrease the accuracy, we don't know.

However, such a decrease of accuracy with the increase of the number of dimensions is not of principal limitations for the proposed method. If the issue becomes critical, the situation can be remedied as follows. When we explained the method using Fig. 1, we made a note that different data sets can be used for estimation of the same *y*-value. In particular, for the data in Fig. 1, it can be done using the datasets $\{Y_1, Y_3, Y_4\}$ or $\{Y_1, Y_2, Y_3\}$. Once we do estimation of the same unknown *y*-value using these datasets, we can use the average value. This average estimation is more accurate, since this time we consider different combinations of data points, whose *y*-values have independent errors. In general, for the independent errors of *y*-values, the error of such an average estimation is proportional to $1/\sqrt{C}$, where *C* is the number of point combinations used for estimation of the same unknown *y*-value. Note that "different combinations of points" include both different reference points and the auxiliary points used for calculation of gradient values.

The same approach simultaneously addresses the possible overfitting, which may occur for certain data types. For instance, an overfitting may happen if the neighboring *y*-values were measured with very different in value errors. Using average estimations of the same *y*-value based on different combinations of predictor points addresses the problem. In many instances, the points-outliers can be detected based on comparison of *y*-values of nearby points. The ones which are significantly greater or smaller than the rest, are good candidates for being outlying points.

An important note has to be made with regard to choosing combinations of points. Since all gradients are tied together via the system of equations, it means that the change of even one point will affect all gradient values. So, we do not to necessarily have to have different points in each combination of input points used for estimation of the same *y*-value. In practical terms, replacing several points for a hundred dimension problem effectively creates a new meaningful *y*-value.

Overall, as we can see, the situations when *y*-values are measured with significant errors can be satisfactory handled in order to reach the required accuracy. This, accordingly, will require additional computations. Computational time will be proportional to the number of point combinations. Since the proposed algorithms have very good computational performance, that will not affect the practicality of methods.

## Discussion

### *Computational performance*

Column 6 in Table 3 shows the calculation time per point for both methods depending on the dimension of space. Calculation performance for the gradient based method can be considered as a good one. Ten times increase in the space dimensionality (from 10 to 100) leads to about 57 times increase in the calculation time. The main source of this increase is the approximately quadratic growth of the number of terms in the system of linear equations, whose solution is required to find the gradients. There are different computational methods and algorithms that allow reducing the calculation time for such tasks, if needed.

We show the range of calculation time for the smooth approximating surface (column 6 in Table 3, the right column). The main consumer of computational resources is the task of finding gradients for each required surrounding point, which is $(N-1)$, where $N$ is the space dimension. Since each point has $(N-1)$ number of $x$-coordinates, it means that if we use the "brute force" approach, we have to calculate $2(N-1)^2$ gradients per each point (the number '2' appeared because we compute gradients in the directions of increase and decrease of $x$-coordinate). This is a huge redundancy, since overwhelmingly we recalculate the same or unnecessary gradients. Recall that in order to find the correction using the smooth approximating function we need only four gradients per each $(x_i, y)$ plane regardless of the space dimension. The maximum calculation time in column 6 in Table 3 corresponds to the case of this maximum redundancy. If we do not calculate unnecessary gradients, using specifics of the problem, then the calculation time can be by orders of magnitude less. In our case, we used the consideration that the points are located on a rectangular mesh. The minimum calculation time presented in the column 6 of the Table 3 corresponds to this scenario. In real problems with irregular non mesh-like location of multidimensional points, an improved computational performance can be achieved by optimization of datasets of points used for finding "left" gradients (see Fig. 3), so that the solution of the appropriate system of linear equations Eqn 7 could produce several needed gradients at once. Overall, this is the area with large reserves of optimization possibilities.

In fact, even with the maximum redundancy, this is a computationally efficient method, given the nature of the problem. This assertion is further enforced by the fact that the nature of the proposed methods makes them very well suited for parallel computing. Such, we can compute gradients for all $N$ points in parallel, since these are independent operations.

Considering computational performance, we did not take into account one contributing factor, which is the search of nearby points, since in our application we generated points on the mesh and then displaced them along all *x*-coordinates. Such search operations are well defined and fast computational procedures. Different algorithms and methods can be used for that purpose, depending on the nature of data, type of their physical storage, structural organization, data and system specifics, etc. By and large, this is rather technical than of principal meaning issue. So, the addition of time required for search procedures should not noticeably affect the overall computational performance of proposed methods.

### *Creating layers for multidimensional y-vectors*

Above, we considered one-dimensional *y*-component. In fact, in practical problems the output may be required to be a multidimensional vector. For instance, in advertising tasks, it could be a vector *y* that has components describing the potential client's purchase power, preference of a car manufacturer, the model, the car color, financing terms, etc. Such a requirement of a multidimensional output can be addressed by the proposed methods using "layers", each of which representing a component of the *y*-vector. In other words, we calculate each *y*-component separately, and then combine them into a single vector *y*, which in this regard can be considered as a vector field. Certainly, we can create the second similar vector field $y_{(2)}$.

Fig. 9 presents these considerations in a graphical form, when from the original dataset *x* first multidimensional the dataset $y_{(1)}$ is created using either the gradient based method or the smooth surface approximation, and then the next multidimensional dataset $y_{(2)}$ is created in a similar way. Note that in terms of number of dimensions, the derived datasets may have smaller, the same or greater dimensions. Such a functionality of the proposed methods provides a great deal of flexibility when working with different kinds of diverse data that have heterogeneous components. The fact that the methods work equally well both with dependent and independent components enhances their universality with regard to the range of problems they can be applied to.

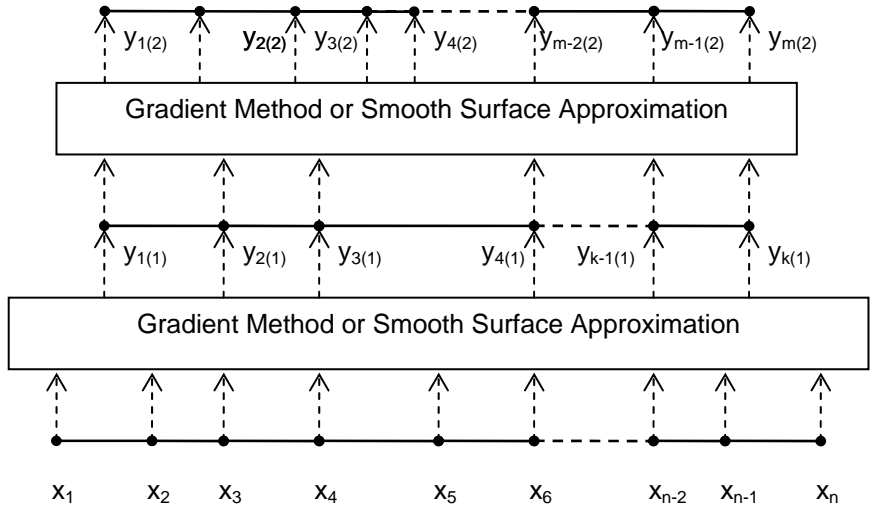

Fig. 9. Creating secondary multidimensional datasets $y_{(1)}$ and $y_{(2)}$ from the original multidimensional dataset $x$. Note that the secondary datasets may have both smaller and greater dimensions.

**Properties of proposed methods with regard to their application**

There are two distinct ways of application of proposed methods. One is when $y$-values do not have errors, or errors are insignificant. There are many practical problems, which fall into this category. Such, in retail and trading applications the price is known precisely. Many measurements in practical terms can be considered as error-free too.

The second approach accounts for possible errors in measuring $y$-values. In this case, if the accuracy is not sufficient, one should calculate the same $y$-value using several combinations of points, as it was discussed, and then use the average.

As is usually the case with the general methods, they can be applied within very different areas, in which this kind of data processing is required. The methods are especially fit for processing of big high dimensional datasets. There is no lack of different methods of data processing presently, especially statistical ones. However, many of them are not used in practical applications. The reasons could be their complexity, when there is not enough qualified staff to use them prudently, difficulty of implementation, like developing software data processing systems, availability of other methods, whose performance might be even subpar for the task, but people are familiar with and know how to work with them.

With regard to these obstacles, impeding the usage of many data processing methods, the gradient based method, in our opinion, falls rather into the category of simple and highly efficient methods, both analytically and from the perspective of software implementation. It also has an excellent, if not outstanding, computational performance compared to other methods, and provides very reasonable accuracy, acceptable in many practical applications. The smaller the discretization intervals, the more accurate results the method produces. Coupled with an excellent computational performance, it means that the gradient based method is especially good for big datasets with relatively small discrete intervals ('small' compared to characteristic lengths of irregularities along the appropriate *x*-coordinates of considered surfaces), or for relatively smooth multidimensional surfaces, which do not have peaks or troughs with sharp edges.

The method with smooth approximating surfaces is presenting rather the next, middle level of complexity, both analytically and from the implementation perspective. With the right application design [7], it can be implemented from scratch in several days. In a production environment, depending on the requirements, implementation of a production application might take several weeks for a team of 2-3 software developers.

The main advantage of the method is its very high accuracy (by 2-3 orders of magnitude compared to the gradient based method, as our results presented in Tables 2 and 3 showed). The method is especially good for the sparse datasets (of the order of characteristic lengths of surface irregularities), although, of course, it works equally well with smaller discretization intervals too, providing an excellent accuracy. The good thing about this method is its high flexibility. We used smooth functions requiring the continuity of first derivatives of the curves presenting surface's cross-sections by (*x*,*y*) planes. Depending on a particular surface, other requirements can be set and other approximating functions can be used. For instance, if the surface is composed of multidimensional hyper-planes (recall "Stealth" aircraft as a 3-D example), then the gradient based method, in fact, produces the most adequate results.

The very idea of using local area for finding *y*-values complies with nature of many real data, since there are lots of real problems, in which correlation between remote points could be weak or inessential for the task. In such situations, creating an entire surface is not needed, and if done, it could be rather waste of resources. The local area approach also provides flexibility with regard to the size of the local area, which can be matched with the task and characteristics of data. Such, for adjusting gradients in case of smooth surfaces, we can use more than four points if more distant points somehow affect the approximating curves.

The computational performance of the method, judging by our results, can be excellent, if the calculation of gradients is properly optimized, as our example showed, although the method works well even without such optimization. Besides, the nature of both methods is such, that the parallel computing can be easily used with them, which could further significantly enhance the computational performance.

A note should be made about the possibility to use these methods in cluster analysis. We considered multidimensional surfaces. However, in the same way, one can use the proposed methods for making decisions, if some entry point belongs to a cluster or not. For that the obtained *y*-value, and possibly the *x*-coordinates, have to be compared against the ranges defining the cluster's boundaries.

Overall, as the authors, we think that both methods have very good prerequisites for their wide usage in practical applications (while remembering that one of the main properties of life is that it can easily override any foresight).

**Possible areas of applicability of the proposed methods**

The most obvious direct application is when one has a dataset of a certain type with known *y*-values (scalar or vector ones), and wants to find the *y*-value for a new entry. In this regard, the *x*-values can represent characteristics of virtually any entity - individual people and groups, animals, recognized targets, economical, environmental, climate, geological, societal conditions, their dynamics, regional specifics, industrial products, production and construction processes, etc. We already mentioned advertising as a particular area of application, while economical, political, sociological information can be processed in a similar way.

Other possible areas of application the methods ask for is coding of information. Sound tracks are naturally fit for this purpose, as well as coding of visual information. For instance, in the last case, one can use three layers of RGB colors to reproduce the picture, although numerous other ways of representing color and location information are possible. In fact, the algorithms present many opportunities for new developments in coding of different types of information. With regard to coding, we can see at least two important advantages of proposed methods. The first is the possibility of variable discretization depending on the coded information. For instance, if an image has a uniform sky color in a large area, there is no need to code this fraction of the image with the same discrete intervals as small flowers on the same image. The second advantage is the possibility to use approximating functions corresponding to the information to be coded.

*Relationship of our algorithms with neural networks*

The next possible area of application might relate to taking care of some functionality presently served by other algorithms and methods, when the proposed methods provide algorithmic or computational advantage. In this regard, artificial neural networks probably should be also considered, although presently many questions, which can be asked, have no answer. However, at least in some aspects, we have a close similarity. Indeed, the smooth surface algorithm also creates a smooth multidimensional surface, as the artificial neural networks algorithm does. Neural networks do this based on certain criteria. Presently, our method rather lacks such strict criteria, but there are no principal limitations for doing this. The issue requires studies, if such a need ever arises. On the other hand, while the neural networks algorithm requires substantial computational resources to create an approximating surface, our algorithms do this instantly, at any required point. The often requested ability of the neural networks algorithm to create "signatures" of objects they describe should be addressed by the proposed algorithms too, if one decides to use them, for instance, in image recognition problems. We did not study whether such an enhancement can be introduced for our methods. However, at a first glance, there is a potential for doing so. The reason for such an assertion is the great number of degrees of freedom, which the new methods introduce and support. For instance, the pace of changing the color or its components on the visual image, the amplitude of such change for different color components can be easily and very quickly obtained using the proposed methods. So, overall, the first impression is that replacing in some steps certain functionalities presently supported by artificial neural networks looks feasible, although, of course, the problem requires intensive studies.

## Conclusion

The idea of inventing new practical methods for multidimensional regression and interpolation in our case originated from the problems of using artificial neural networks for certain tasks, first of all for highly irregular surfaces, and especially for the ones with sharp multidimensional peaks, for which the artificial neural networks did not provide solutions. In order to address such problems, we came with a new concept and first introduced the method based solely on using gradients, and then its enhancement - the second method that is based on smooth approximating functions. As our numerical examples show, both methods provide very good functional and computational performance, with the second method offering by 3-4 orders of magnitude better accuracy. This is something that no ad hoc method and minor improvement usually allow to achieve. So, what is the source of such high efficiency?

**The relationship of proposed methods and statistical approaches**

The proposed methods have many advantages compared to known approaches. Some of them were discussed, while some were left outside. The thing is that we proposed new methods, whose conceptual foundation substantially differs from traditional paradigms, especially statistical ones. Statistical methods generally assume usage of a statistical model, to which the correspondence of data is compared against based on certain (supposedly typical statistical) criterion, like the loss or likelihood functions.

Such largely artificial dedication of the present statistics to a fairly restricted collection of fundamental concepts imposes rather unnecessary limitations and restrictiveness on the discipline, thus impeding the diversity of its development. This misbalance became especially noticeable in the last decade, when the quickly growing availability of "big data" encountered a certain lack of efficient genially statistical methods for data processing. Artificial neural networks (ANN), indeed, helped to fill this demand. However, by conventional criterion, ANNs are rather not statistical methods, but an interpolation computational technique. It is good that not only computer science, but many statistical departments at universities included ANNs into their studies; it created the precedence for the eventual expansion of statistics beyond its present boundaries. However, the fact is that ANNs largely originated from the studies in computer science, not in statistics, which would be impossible in the present state of the discipline.

By the same token, the concepts and the methods that we proposed belong to the realm of computational methods. The methods use and can be further enhanced with statistical notions, but at the core these are computational methods. This specific "stand aloneness" from the typically statistical methods should be understood in order to fully use the potential and advantages these methods offer. Attempts to impose classical statistical approaches on them without accounting for their computational nature or, worse, squeezing them into the shell of statistical methods, will unlikely do any good for their efficiency and performance. Such improvements would be more appropriate using less restrictive and simpler new ideas and concepts, since the methods provide many degrees of freedom and lots of space for doing this.

**Methods' limitations versus the informational data potential**

During the study, it became more and more obvious that the proposed methods do have very wide range of applicability, are highly accurate, and have an outstanding functional and computational performance. These are not the features of ad hoc methods, but rather the properties of new conceptual breakthroughs, however pretentious these words may sound at first. In the work [8], one of the authors showed the origin of limitations on the probability of recognition in recognition problems. Although such limitations used to be considered of fundamental nature, in fact, that was the limitation of mathematical methods, and in particular the ones using likelihood functions. The amount of input information was sufficient to provide much better accuracy of recognition; it was a matter of finding the right mathematical interpretation method, which could allow extracting the needed information.

In many instances, it is possible to make evaluation (qualitative or/and quantitative) of potential information capabilities contained in data, as it was the case with the problem in [8]. To a certain extent, a similar situation occurred with inability to use artificial neural networks for sharp multidimensional peaks. The input information seemed as a sufficient one for the task, but the existing method did not allow doing this. When such a situation occurs, the best remedy would be introducing a new method, better off a new method based on a new general concept. This is how we view the proposed methods. Indeed, the introduced concept allows for many further developments; few of them were discussed, while many others were not. The novelty of the introduced paradigm may mask these numerous possibilities, but they are there, and lots of them.

## Acknowledgements

The authors thank K. Y. Shestopaloff for the help with literature review.